\documentstyle[aps,twocolumn,epsf]{revtex}
\begin{document}
\bibliographystyle{prsty}
\title{  Calculation of ionization within the close-coupling formalism }
\draft

\author{Andris T. Stelbovics}

\address{
Centre for Atomic, Molecular and Surface Physics,
School of Mathematical and Physical Sciences,
Murdoch University,
Perth 6150, Australia
}

\date{\today}

\maketitle


\begin{abstract}

An adaption of the convergent close--coupling method (CCC) to calculation of differential ionization cross sections
 [I. Bray and D. V. Fursa, Phys. Rev. A {\bf 54}, 2991 (1996)] 
is analyzed in the context of the Temkin-Poet model. 
The asymptotic scattering wave functions and the unitarity relation are given for the the model.
Our analysis is
used to  reconcile the recent criticism by G. Bencze and C. Chandler, [Phys. Rev. A {\bf 59}, 3129 (1999)] 
of the formulation of Bray and Fursa.
It is concluded the use  of ``distinguishable'' electrons as proposed by Bray and Fursa is not essential
in the close-coupling description of  ionization. 
\end{abstract}
\pacs{34.80.Bm, 34.80.Dp, 03.65.Nk}
\narrowtext

The CCC method was developed initially \cite{BS92,BS92l} to describe scattering of electrons from one-electron targets. The method made use of
Laguerre basis states that are are $L^2$ functions to discretize the target continuum. The method provided convergent amplitudes for scattering 
to low--lying discrete states and total--ionization cross--sections. The theory is fully symmetrized since it is based on 
a symmetrized expansion of the two--electron wave function. 
Bray and Fursa \cite{BF96} used the fact that the positive--energy pseudostates were approximations to true continuum functions 
to propose a method for calculating energy differential cross sections for ionization. 
They did this in  a way that preserved ``two--particle'' unitarity implicit in the CCC
formalism with the  $L^2$ target states. This had the consequence of yielding different magnitude amplitudes $C^S(\epsilon_a,\epsilon_b)$ and  
$C^S(\epsilon_b,\epsilon_a)$ for the same physically indistinguishable process of  the two ionized electrons emerging with energies 
$\epsilon_a,\epsilon_b$, depending on which electron, $a$ or $b$, is represented by a Coulomb wave. Bray and Fursa argued that this asymmetry
could be interpreted in terms of ``distinguishable'' electrons.
This has been criticized by Bencze and Chandler \cite{BC99} (BC) who argue that the
symmetrization property of the amplitudes is a fundamental tenet of scattering theory and 
that the CCC amplitudes must satisfy this property
counter to the numerical evidence. They
conclude that ``The numerical CCC amplitudes have not, therefore, converged to accurate approximations of the exact amplitudes.'' 

It is the dichotomy  between a method of calculating ionization processes that seems very impressive in the quality of agreement achieved
between theory and experiment, and the criticism of the model on a very fundamental level, that provides the motivation for the present work. 
To make the discussion as transparent as possible we choose the Temkin--Poet 
model (TP) \cite{T62,P78} of electron--hydrogen scattering. This model is the solution of the scattering problem assuming spherical averaging
over both electrons and solving for the total angular momentum zero two--electron wave function. The Schr\"odinger equation becomes

\begin{equation}
(\frac{\partial^2}{\partial x^2}+\frac{\partial^2}{\partial y^2}+\frac{2}{min(x,y)}+2E) \Psi^S(x,y) =0, \label{se}
\end{equation}

with boundary and symmetry  conditions
\begin{eqnarray}
\Psi^S(x,0)= \Psi^S(0,y)=0, \label{bc} \\
\Psi^S(y,x)= (-1)^S\Psi^S(x,y). \label{sc}
\end{eqnarray}

Since Eq. (\ref{se}) is separable in the regions $x>y$, and $x<y$, the solutions satisfying physical boundary conditions may be written as

\begin{eqnarray}
\Psi^S(x,y)= u_i(y)e^{-ik_ix}+\sum_{j=1}^{\infty} C_{ji}^{S}u_j(y)e^{+ik_jx}  \nonumber \\
+\int_{0}^{\infty}d\epsilon_b C^{S}(\epsilon_a,\epsilon_b)_iu^{-}_{\epsilon_b}(y)e^{+ik_ax}, x>y. \label{wf_exact}
\end{eqnarray}
The solution is extended to the region  $x<y$ by using the symmetry condition (\ref{sc}). The solution (\ref{wf_exact}) has been written assuming the total energy 
is above the ionization threshold so the momenta are fixed according to
$E=\epsilon_i+\frac{1}{2}k^2_i=\epsilon_j+\frac{1}{2}k^2_j=\epsilon_a+\epsilon_b.$
As usual the momenta for the ionized electrons are given by $ \frac{1}{2}k^2_{a(b)}=\epsilon_{a(b)}$. The $u_j$ are the manifold of 
bound target states of the hydrogen atom target with $l=0$ angular momentum. The continuum functions are the regular $l=0$ 
Coulomb functions
multipled by a Coulomb phase and
normalized to a delta function in the energy. It is obvious from the form of the solution that the $C^S$ coefficients are related to
the $S$--matrix elements since we have written the expansion in terms of incoming and outgoing  waves. We are not concerned here with the 
problem of solving for these coefficients; this problem has been addressed in \cite{T62,P78}. 
Rather we wish to establish the form of the asymptotic scattering functions and the form of unitarity
relation that one obtains for this model when the ionization channel is open. 
For scattering to discrete states
the asymptotic form of Equ. (\ref{wf_exact}) is
  
\begin{equation}
\Psi^S_{disc}(x,y) \stackrel{\sim}{_{x \rightarrow \infty }} \sum_{j=1}^{\infty} C_{ji}^{S}u_j(y)e^{+ik_jx}. \label{psi_disc}
\end{equation}

The asymptotic form for the ionization requires that both particles go to infinity and that each is represented by an outgoing wave. Therefore it is 
necessary to extract these forms from the integral term in (\ref{wf_exact}). 
Normalizing the $u^{-}_{\epsilon_b}$ to a delta function in energy asymptotically
\begin{equation}
u^{-}_{\epsilon_b}(y) \stackrel{\sim}{_{y \rightarrow \infty}}\sqrt{\frac{2}{\pi k_b}} e^{-i\delta _c} \
\sin (k_by + \frac{1}{k_b} \ln 2k_by + \delta_c). \label{coul}
\end{equation}
The  integral can be readily evaluated as $x,y\rightarrow \infty$ using stationary phase methods (a very readable exposition is given in ref. \cite{BH86}).
It is 
determined by the integrand in the vicinity of the stationary points (see for example \cite{RS65,P77}) of the terms $e^{i k_a x \pm i k_b y}$. 
One finds there is no stationary point for the exponential with $-ik_by$, and hence no contribution to the asymptotic form. The only
contribution comes, as expected, from the outgoing wave giving

\begin{eqnarray}
\Psi^S_{ion}(x,y) \sim C^S(\epsilon_a,\epsilon_b)_ie^{i\pi/4}(2E)^{\frac{1}{4}}\cos \alpha r^{-1/2}  \nonumber \\
\times e^{i(\sqrt {2E} r+\frac{1}{k_b} \ln 2k_by )}, \ \ x>y. \label{psi_ion}
\end{eqnarray}
Here $r=(x^2+y^2)^{\frac{1}{2}}$ and the energies $\epsilon_{a,b}$ are fixed by the stationary point according to

\begin{equation}
\tan \alpha = \frac{y}{x} = \frac{k_b}{k_a},\ \ 
0<\alpha<\frac{\pi}{4}. \label{sp}
\end{equation}
The discrete scattering sector corresponds to $\alpha=0$.

We now derive the unitarity relation using the conservation of flux, incoming and outgoing, 
through a closed surface outside which the asymptotic forms for the wave function are appropriate. 
The current for the TP model is defined as ${\bf j}(x,y)=\Im \Psi ^{S*}\nabla  \Psi ^S$.
The surface we choose to apply the flux conservation law to is shown in Fig.1 and is formed by  $x=0$, $y=0$, $x=a$ and $y=a$. $a$ is assumed to
be large enough to ensure use of the asymptotic form for ionization is justified. It is also useful to define two strip regions running along the
$x$-- \ and $y$--axes bounded by  $x=a^{\eta}$, $y=a^{\eta}$ where $\frac{1}{2}<\eta<1$. It is assumed that the flux for the discrete target state 
wave functions is contained within these strips for a finite number of the states which we may choose to be as large as we wish. 
Conservation of flux requires that
$\int_{\Box}{\bf j.}d{\bf n}=0$.
On account of the boundary conditions (\ref{bc}) and symmetry condition (\ref{sc}) this reduces to 
\begin{equation}
\int_{0}^{a}{j_x(x=a,y)}dy=0.
\end{equation}
The incoming incident and the discrete outgoing currents are  confined to the strips and as such we see  that
\begin{eqnarray}
f_{(inc)}=-k_i\int_{0}^{a^{\eta}}|u_i(y)|^2dy \rightarrow -k_i, \ \ \rm {as} \ \  a\rightarrow\infty,\\
f_{(disc)}\rightarrow+\sum_{j=1}^{\infty}k_j|C^S_{ji}|^2 \ \ \rm {as} \ \  a\rightarrow\infty.
\end{eqnarray}
Calculation of the ionization flux requires  more care because the asymptotic form of the Coulomb wave (\ref{coul}) is valid
only when $k_by>>1$.
Thus in order to calculate the ionization flux we have to ensure that this condition is met throughout the interval of integration $[a^{\eta},a]$.
 (We
note that there is a minimum value of $k_b$ fixed by the stationary phase condition (\ref{sp}).) These constraints
may be satisfied by choosing

\begin{equation}
\frac{1}{2}<\eta<1, \\
 \ \ a>>(2E)^{-\frac{1}{2}}. 
\end{equation}
The total ionization flux using the current from Eq. (\ref{psi_ion}), is in the limit of large a
\begin{equation}
f_{(ion)} = (2E)^{\frac{3}{2}}\int _{a^{\eta}}^{a} |C^S(\epsilon _a,\epsilon _b)_i|^2 \frac{a^2}{(a^2+y^2)^{\frac{3}{2}}}dy.
\end{equation}
The remaining terms in the current involve overlaps between ionization and discrete wave functions and are readily shown to 
contribute vanishing flux in the limit of large a.
Collecting all the flux contributions the required unitarity relation in the limit $a\rightarrow\infty$ is
\begin{equation}
1=\sum_{j=1}^{\infty}\frac{k_j}{k_i}|C^S_{ji}|^2+\int _{0}^{\frac{E}{2}} |C^S(\epsilon _a,\epsilon _b)_i|^2\frac{k_a^2}{k_bk_i}d\epsilon _b. \label{u}
\end{equation}
Thus the ionization contribution to unitarity requires $C^S(\epsilon _a,\epsilon _b)_i$ only for $0<\epsilon _b \le \frac{E}{2}$. 

Let us now turn to the close--coupling (CC) approach and its description of ionization. One begins with an explicitly symmetrized wave function:
\begin{eqnarray}
\Psi^{S(CC)}(x,y)= (1+(-1)^SP(x,y))(\sum_{j=1}^{\infty} f_{ji}^{S}(x)u_j(y) \nonumber \\
+\int_{0}^{\infty}d\epsilon_b f^{S}(\epsilon_a,\epsilon_b)_i(x)u_{\epsilon_b}(y)) \label{psi_cc}
\end{eqnarray}
where $P$ is the permutation operator for $x,y$ and 
\begin{eqnarray}
f^S_{ji}\stackrel{\sim}{_{x \rightarrow \infty }}\delta _{ij}e^{-ik_ix}+C^{S(CC)}_{ji}e^{+ik_jx}, \\
f^{S(CC)}(\epsilon_a,\epsilon_b)_i(x)\stackrel{\sim}{_{x \rightarrow \infty }}C^{S(CC)}(\epsilon_a,\epsilon_b)_ie^{+ik_ax}. \label{f_cc}
\end{eqnarray}
The asymptotic form of the CC wave function can be found using the methods already discussed. It is readily seen that the discrete scattering
boundary condition (\ref{psi_disc}) is unaltered but the ionization asymptotic form becomes
\begin{eqnarray}
\Psi^{S(CC)}_{ion}(x,y) \sim e^{i\pi/4}(2E)^{\frac{1}{4}}\cos \alpha r^{-1/2} e^{i\sqrt {2E} r} \nonumber \\
\times (e^{i\frac{1}{k_b} \ln 2k_by} C^{S(CC)}(\epsilon_a,\epsilon_b)  \nonumber \\
+ (-1)^Se^{i\frac{1}{k_a} \ln 2k_ax}C^{S(CC)}(\epsilon_b,\epsilon_a)). \label{psi_symm_ion}
\end{eqnarray}
This form has the unsatisfactory feature that it cannot be expressed in the form of an outgoing wave multipled by an amplitude that depends only on
$\epsilon_a,\epsilon_b$ due to the logarithmic phases. One is forced to conclude that the symmetrization, on its own, does not lead to a
generally  acceptable form for ionization. 
If the long--range logarithmic phase were not present, the asymptotic form is perfectly acceptable since one may write
\begin{eqnarray}
 C^{S(CC)}_{sym}(\epsilon_a,\epsilon_b)=C^{S(CC)}(\epsilon_a,\epsilon_b) 
+ (-1)^SC^{S(CC)}(\epsilon_b,\epsilon_a)  \label{cc_sym}
\end{eqnarray}
and $ C^{S(CC)}_{sym}(\epsilon_a,\epsilon_b)=(-1)^S C^{S(CC)}_{sym}(\epsilon_b,\epsilon_a)$ as demanded by conventional quantum theory for
identical particles as emphasised by Bencze and Chandler \cite{BC99}. 
Further, note also that there is no unique form for the $f^{S(CC)}( \epsilon_a,\epsilon_b)$ in Eq. (\ref{psi_cc}), and hence for the
$C^{S(CC)}( \epsilon_a,\epsilon_b)$ of Eq. (\ref{f_cc}). Thus the $N\rightarrow\infty$ limit in BC cannot be taken as unambiguously as they assume
and the CCC equations do not have to converge necessarily to a symmetrized ionization amplitude.

Let us now examine how to apply these considerations to the CCC method for ionization. In this method the expansion over the complete 
set of target states is replaced by a
set of $L^2$ target states generated by diagonalizing the Hamiltonian in an $N$--function subspace  of the Laguerre basis. The limit $N\rightarrow \infty$
gives the CC expansion (\ref{psi_cc}).
The asymptotic form of the wave function is
\begin{eqnarray}
\Psi^{S(N)}_{disc}(x,y) \stackrel{\sim}{_{x \rightarrow \infty }} u_i^{(N)}(y)e^{-ik_i^{(N)}x}+ \nonumber \\
+\sum_{j=1}^{N} \ C_{ji}^{S(N)}u_j^{(N)}(y) e^{+ik_j^{(N)}x}. \label{psi_cc_disc}
\end{eqnarray}
The $u_j^{(N)}$ are the $L^2$ approximate target states and their energies 
are $\epsilon_{j}^{(N)}$. Because the  $u_j^{(N)}$ are $L^2$ states the scattering flux for all the states is confined to the strip regions in Fig 1.
Thus as it stands there is no flux in the ionization sector $0<\alpha<\frac{\pi}{2}$. 
In order to sensibly extract an ionization amplitude from Eq. (\ref{psi_cc_disc}) one must make an approximation that yields a true continuum function for the 
positive--energy 
target states. The following is an obvious possibility:  insert the completeness relation for the true target states in the variable $y$ and 
let $y\rightarrow\infty$. Then one may write
\begin{eqnarray}
\Psi^{S(N)}_{ion}(x,y) \stackrel{\sim}{_{{x \rightarrow \infty},y \rightarrow \infty}}\int_{0}^{E}d\epsilon_b \
(1+(-1)^SP(x,y)) \nonumber \\
\times \sum_{j:0<\epsilon_j^{(N)}<E}C^{S(N)}_{ji}\langle u_{\epsilon_b}|u_j^{(N)} \rangle u_{\epsilon_b}(y)  e^{+ik_ax},
\end{eqnarray}
where use was made of the fact that $\langle u_{\epsilon_b}|u_j^{(N)} \rangle$ is highly peaked about $\epsilon_j ^{(N)}$ 
and energy conservation to
replace $ e^{+ik_j^{(N)}x}$ by  $ e^{+ik_ax}$. 
It is straightforward to prove this leads to the expression  of Bray and Fursa \cite{BF96}:
\begin{equation}
C^S(\frac{1}{2}{k^{(N)}_j}^2, \epsilon _j^{(N)})_i\sim C^{S(N)}_{ji}\langle u_{\epsilon_j^{(N)}}|u_j^{(N)} \rangle.
\end{equation}

These amplitudes may be interpolated to provide the (unsymmetric) $C^{S(N)(CC)}(\epsilon_a,\epsilon_b)$ which are then used to construct
the symmetrized amplitude (\ref{cc_sym}). It is this  amplitude that must be used in the unitarity relation (\ref{u}).
Bray and Fursa \cite{BF96} one the other hand argued that the unsymmetric amplitudes must be summed incoherently 
to satisfy unitarity, a
suggestion that is equivalent to treating the two electrons as distinguishable
and the integration is performed over the interval  $0\le\epsilon_b\le E$. 
Bray further suggested \cite{B97l} that in the limit 
$N\rightarrow \infty$  the CCC(N) amplitudes 
for $\frac{E}{2}<\epsilon_b\le E$ would converge to zero.
This has also been criticized by BC and rebutted by Bray
\cite{B99}. 
We now give a specific example where the 
distinguishable--electron hypothesis can be compared directly with coherent summation of identical-particle amplitudes.

Consider the letter of Bray \cite{B97l} that reports  CCC(N) calculations for the TP singlet scattering model at E=2Ryd. It is
stated there that `` there is a clear lack of convergence in going from 10 to 50 states''. 
However, curiously, on inspection of  his Fig. 2  \cite{B97l} there appears to be some convergence in the region of  $\frac{E}{2}$.
We can use the fact that the CCC(N) cross sections are convergent there to
extract a sequence of interpolated CCC(N) values for the  $\frac{E}{2}$ cross section.
In Table 1 we show the results obtained by a spline interpolation. 
The convergence is apparent. The reason this energy is critical is that it is the one energy where the symmetrized CC amplitude (\ref{cc_sym})
properly satisfies the boundary conditions of Eq. (\ref{psi_ion}). Thus the symmetrized amplitude leads directly
to a cross section proportional to $4|C^{0CCC(N)}(\frac{E}{2},\frac{E}{2})|^2$.  In the table are also shown the results of a very recent calculation
by Baertschy Rescigno and McCurdy \cite{BRM99} using an exterior complex scaling method \cite{MRB97} that circumvents the problem of matching to Coulomb
ionization boundary conditions. The result through use of the correctly symmetrized CC amplitude is in 
good agreement with their value. The symmetrized
CC amplitude of course  is  symmetric about $\frac{E}{2}$ and no step occurs at  $\frac{E}{2}$. As we move away from this energy 
on the assumption that the neglect of logarithmic phase factors is a secondary effect, the cross section should move smoothly 
up from its minimum value. This is the behavior
observed by  Baertschy et al \cite{BRM99}. 

Interestingly the the total ionization cross section obtained by Baertschy et al is in excellent
agreement with the CCC predictions \cite {BS94atd}. Using the symmetrized amplitude and integrating up to  $\frac{E}{2}$
seems to lead to a very similar answer as using incoherent summation and integrating from  $ 0\le\epsilon_b\le E$. 
Thus all the features of the TP model are explained by CCC theory provided one uses
a fully symmetrized ionization amplitude. 
In case it is argued that the  example chosen is fortuitous to our case,
we  also analyzed the TP cross-section data presented by Bartschat and Bray 
\cite{BB96praR} at  E=3Ryd. The results are shown in Table 1.
Again there is convergence to the value of Baertschy et al.

Our conclusions are as follows. For CC methods in general, the symmetrized expansion gives rise to ionization scattering wave function asymptotic forms
that cannot be simply expressed as outgoing wave in hyperspherical coordinates multiplied by an amplitude unless there are special circumstances. For
TP triplet scattering one such special case arises in that CCC calculations yield an unsymmetric amplitude 
that converges to zero for $\epsilon_b >\frac{E}{2}$. 
In this case a symmetrized amplitude can be made trivially.  
Thus if the TP CCC triplet scattering calculations  converge they must converge to the correct model values and this is what is observed \cite{BRM99}. 
For TP singlet scattering at the equal--energy--sharing kinematics,  $\epsilon_b =\epsilon_a=\frac{E}{2}$,  the symmetrized CC  amplitude 
satisfies the correct ionization asymptotic form including logarithmic phases.
Using the  converged unsymmetric singlet CCC amplitudes to construct the symmetrized CC cross section  
we get very good  agreement with an independent method of solution for the Temkin-Poet model. 
These kinematics are ones  where coherence of amplitudes is maximal. The correct symmetrized singlet cross section is therefore
a factor of four larger than the value from the raw CCC calculations. For asymmetric energy kinematics the symmetrized amplitude
will be very similar to the published CCC calculations below  $\frac{E}{2}$ because the unsymmetric amplitudes above $\frac{E}{2}$ are small. 

We make the following observation regarding Bray's step hypothesis. It is classical Fourier theory behaviour
that Fourier series which represent functions with steps, converge, at the point of discontinuity, 
to the midpoint of the step. 
If the singlet CCC amplitude has such a step, one might expect the amplitude  to converge to the midpoint at the step 
and the cross-section therefore to one quarter of its step height. This
is consistent with our findings.

The conclusions regarding coherent summation of amplitudes may carry over to the full e--H and e--He  problems. 
For example, it has been found that the CCC results, after incoherent combination, at
equal--energy--sharing are typically a factor of two too low, i.e. a factor of four on the raw CCC results \cite {BFRE97_64}.
For asymmetric kinematics it is likely that the CCC converges to the true amplitudes contrary to the analysis of BC.

The author is indebted to Igor Bray, for stimulating discussions and points of clarification regarding his work and
for providing access to his data base of CCC calculations. The support of the Australian Research Council and Murdoch University is also
acknowledged.


\begin{table}
\caption{Singlet energy DCS (in units of $\pi a_o^2$ per Ryd) of
the TP e--H scattering problem 
are shown at a total three--body energy of 2Ryd \protect\cite{B97l} and 3Ryd
\protect\cite{BB96praR} for  CCC(N) calculations 
with the kinematics $\epsilon_b=\epsilon_a=\frac{E}{2}$. They
are multipled by a factor of
4 for reasons given in the text. $N_{pos}$ is the number of positive energy open-channel
pseudostates in each calculation. The CCC(N) numbers are compared with the calculation
of Baertschy, Rescigno and McCurdy \protect\cite{BRM99}.
}
\begin{center}
\begin{tabular}{cccccc} 
$E(Ryd)$ & 
        \multicolumn{4}{c}{ $4CCC(N,N_{pos})$}
        &  $BRM$ \\ \hline
& (20,8) & (30,13) & (40,18) & (50,24) &  \\
$2$& 0.0172 & 0.0145 & 0.0137& 0.0135 &0.0140  \\ \hline
& (42,19) & (44,20) & (46,22) & (48,23) &   \\
$3$& 0.0072 & 0.0081 & 0.0069& 0.0069 &0.0068  \\
\end{tabular}
\end{center}
\end{table}
\begin{figure}
\epsfxsize=10truecm\epsffile{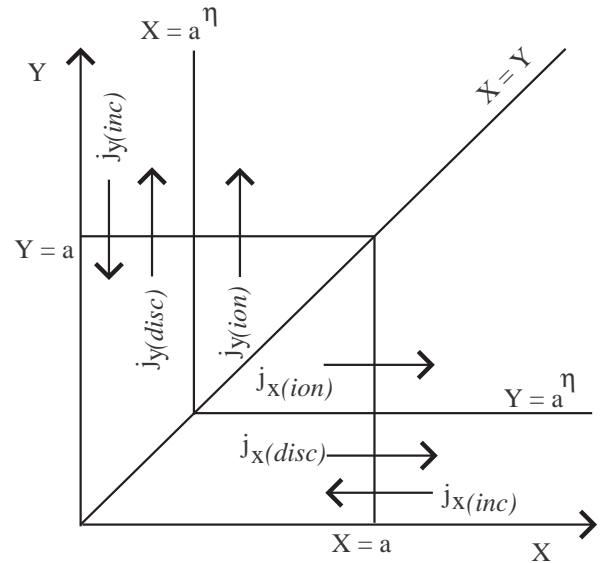}
\caption{
The figure shows the closed rectangular surface 
and incoming, discrete and ionization currents that
pass through the surface. 
See the text for details.
}
\end{figure}

\end{document}